\documentclass[aps,prb,twocolumn,superscriptaddress,showpacs]{revtex4-2}
\usepackage{color}
\usepackage{graphicx}% Include figure files
\usepackage{dcolumn}% Align table columns on decimal point
\usepackage{bm}% bold math
\usepackage[mathlines]{lineno}% Enable numbering of text and display math
\usepackage[colorlinks,linkcolor=blue,anchorcolor=blue,citecolor=blue,urlcolor=blue,hyperindex,CJKbookmarks,dvipdfm]{hyperref}

\usepackage{amsmath}
\usepackage{txfonts}

\usepackage{tabularx}
\usepackage{array}
\usepackage{footnote}
\usepackage{dcolumn}% Align table columns on decimal point
\usepackage{booktabs}

\begin{document}

\renewcommand\arraystretch{1.1} %used to adjust the height of a table, need array package.
\newcolumntype{L}[1]{>{\raggedright\arraybackslash}p{#1}}
\newcolumntype{C}[1]{>{\centering\arraybackslash}p{#1}}
\newcolumntype{R}[1]{>{\raggedleft\arraybackslash}p{#1}}

\title{$\sigma$ bands driven high-temperature superconductivity in hydrogenated hexagonal BC$_3$ monolayer}

\author{Guo Chen}\affiliation{Department of Physics, School of Physical Science and Technology, Ningbo University, Zhejiang 315211, China}
\author{Ru Zheng}\affiliation{Department of Physics, School of Physical Science and Technology, Ningbo University, Zhejiang 315211, China}
\author{Jin-Hua Sun}\affiliation{Department of Physics, School of Physical Science and Technology, Ningbo University, Zhejiang 315211, China}
\author{Fengjie Ma}\affiliation{The Center for Advanced Quantum Studies and Department of Physics, Beijing Normal University, Beijing 100875, China}
\author{Xun-Wang Yan}\affiliation{College of Physics and Engineering, Qufu Normal University, Shandong 273165, China}
\author{Miao Gao}\email{gaomiao@nbu.edu.cn}\affiliation{Department of Physics, School of Physical Science and Technology, Ningbo University, Zhejiang 315211, China}
\author{Tian Cui}\email{cuitian@nbu.edu.cn}\affiliation{Department of Physics, School of Physical Science and Technology, Ningbo University, Zhejiang 315211, China}%\email{cuitian@nbu.edu.cn}
\author{Zhong-Yi Lu}\email{zlu@ruc.edu.cn}\affiliation{Department of Physics, Renmin University of China, Beijing 100872, China}%\email{zlu@ruc.edu.cn}

\date{\today}

\begin{abstract}
Material with metallic $\sigma$-bonding bands is expected to be a high-temperature superconductor, due to the sensitivity of $\sigma$ electrons to lattice vibration. Based on the first-principles calculations, electronic structures of hydrogenated BC$_3$ monolayers (H$_n$-B$_2$C$_6$ with $n$=1-8) are systematically investigated. At high coverage of hydrogen, the monolayer stabilizes in chair-like $sp^3$-hybridized configurations, leading to the metallization of $\sigma$ bands, especially in H$_7$-B$_2$C$_6$ and H$_8$-B$_2$C$_6$. This metallicity originates from the electron deficiency of boron, compared with insulating graphane. Utilizing Wannier interpolation, the electron-phonon coupling strengths for metallic phases of H$_n$-B$_2$C$_6$ are determined. As expected, strong couplings are identified between the conducting $\sigma$ electrons and low-frequency phonon modes. By solving the anisotropic Eliashberg equations, we confirm that H$_7$-B$_2$C$_6$ and H$_8$-B$_2$C$_6$ are single-gap superconductors with critical temperature being 87 K, exceeding the boiling point of liquid nitrogen. Considering that monolayer BC$_3$ has been synthesized in experiment, our results demonstrate that hydrogenation of two-dimensional BC$_3$ provides a viable pathway to achieve high-temperature superconductivity at ambient pressure.
\end{abstract}

\maketitle

\section{Introduction}
Achieving superconductivity above the boiling point of liquid nitrogen (77 K), remains a paramount challenge in condensed matter physics and materials science, which can dramatically reduce the operational costs of widespread applications of superconductivity in energy transmission, quantum computing, and high-field magnets \cite{Souc-SST28,Larbalestier-NM13,Devoret-Science339,Place-NC12,Larbalestier-Nature414,Sumption-SST34}. Traditionally, high transition temperature ($T_c$) superconductivity has been associated with complex correlated systems such as cuprates \cite{Bednorz-ZPB64,Schilling-Nature363}, iron pnictides/chalcogenides \cite{Kamihara-JACS130,Ren-CPL25,Wang-CPL29}, and nickel-based superconductors \cite{Sun-Nature621,Hou-CPL40,Zhang-NatPhy20}. However, recent advances highlight an alternative route: harnessing hydrogen to realize high-$T_c$ superconductivity within the conventional Bardeen-Cooper-Schrieffer (BCS) mechanism \cite{Bardeen-PR108}.

The singular influence of hydrogen mainly stems from its minimal atomic mass and unique bonding characteristics. It introduces high-frequency optical phonon modes, thereby elevating the energy scale in the superconducting pairing interaction \cite{Ashcroft-PRL21,Ashcroft-PRL92}. The outstanding progresses are demonstrated in extremely pressurized hydrogen-rich compounds, such as H$_3$S \cite{Duan-SciRep4,Drozdov-Nature525}, LaH$_{10}$ \cite{Liu-PNAS114,Peng-PRL119,Drozdov-Nature569,Hong-CPL37}, and LaSc$_2$H$_{24}$ \cite{He-PNAS121,Song-arXiv2510}, whereas hydrogen networks under immense compression exhibit exceptionally strong electron-phonon coupling (EPC), driving $T_c$ to record highs exceeding 200 K and proving that conventional mechanisms can indeed support near-room-temperature superconductivity. However, the necessity for megabar pressures presents a formidable barrier to scalable applications. Potential high-$T_c$ hydrogen-rich superconductors near the ambient pressure have been suggested by theoretical calculations \cite{Gao-PRB104,Gao-PRB107,Dolui-PRL132,Ouyang-PRB111,Wan-PRB112}. Currently, phonon-mediated superconductivity with $T_c$ above the liquid-nitrogen temperature has not yet been realized.

A highly attractive alternative is the incorporation of hydrogen into low-dimensional materials. Dimensional confinement itself can enhance the electronic density of states (DOS) at the Fermi level, i.e. $N(0)$, and when coupled with hydrogenation, it creates a synergistic platform for high-$T_c$ superconductivity at ambient pressure. Pioneering theoretical work identified hole-doped graphane, hydrogenated graphene, as a potential high-$T_c$ superconductor, highlighting $sp^3$-bonded diamond-like two-dimensional (2D) structure could yield a giant EPC. Using rigid-band model, it was predicted that $T_c$ can reach 90 K with hole doping level achieving 6\%-10\% \cite{Savini-PRL105}. This concept has been validated and expanded. For instance, adsorbing hydrogen onto monolayer MgB$_2$ was shown not only to raise substantially but also to create new, strongly coupled phonon modes and electronic states \cite{Bekaert-PRL123}. Subsequent predictions have revealed a rich family of hydrogenated 2D materials, such as Janus MoSH monolayer \cite{Liu-PRB105}, HCaB$_3$ \cite{Yang-CPL40}, hydrogenated metal diborides $M_2$B$_2$ and monolayer lithium borocarbides \cite{Han-PRM7,Liu-PRR6}, which exhibit robust multi-gap superconductivity. These predictions still await further experimental verification.

Although the predicted high-$T_c$ superconductivity in hole-doped graphane is absorbing, finding suitable precursors for synthesizing hole-doped graphane-like structures is crucial for experimental realization. Notably, it was suggested that replacing certain content of carbon with boron is a feasible way to introduce holes into graphane \cite{Savini-PRL105}. Thus, starting directly from the B$_x$C$_{1-x}$ monolayer structure and subjecting it to hydrogenation may represent a reasonable path. Indeed, the synthesis of graphite material BC$_3$, via the interaction of benzene and boron trichloride, was reported in 1986 \cite{Kouvetakis-JChemSoc1986}. An uniform BC$_3$ monolayer with excellent crystalline quality was successfully grown in an epitaxial way on the NbB$_2$(0001) substrate \cite{Yanagisawa-PRL93,Tanaka-SSC136,Yanagisawa-PRB73}. Monolayer BC$_3$ is an indirect semiconductor with a band gap of 0.66 eV \cite{Tomanek-PRB37}, doping holes in $ABC$-stacked BC$_3$ may induce MgB$_2$-like superconductivity with optimal $T_c$ being 22 K \cite{Ribeiro-PRB69}. In particular, full hydrogenation induces a transition from semiconducting to metallic behavior in monolayer BC$_3$ \cite{Ding-JPCC113,Chuang-Nanotechnology22}. Therefore, fully hydrogenated monolayer BC$_3$ can be regarded as a heavily hole-doped graphane, with doping concentration extended from 10\% to 25\%. The EPC and potential phonon-mediated superconductivity in hydrogenated monolayer BC$_3$ are desired to be explored.

In this work, we systematically investigate the atomic and electronic structures of hydrogenated BC$_3$ monolayer, based on the first-principles density functional theory calculations. For metallic phases, the lattice dynamics and EPC are accurately computed to ascertain whether there exists a superconducting transition or not. For clarity, the compounds are named as H$_n$-B$_2$C$_6$, since there are two boron atoms and six carbon atoms in the unit cell, with $n$ representing the coverage of hydrogen. Following hydrogen adsorption, the planar structure of BC$_3$ begins to curl, undergoing a transition from $sp^2$ to $sp^3$ hybridization. For hydrogen coverages corresponding to $n$=1, 3, 5, 7, and 8, H$_n$-B$_2$C$_6$ exhibit metallic characteristics. Notably, by solving the anisotropic Eliashberg equations, the superconducting $T_c$ is calculated to be 87 K for both $n$=7 and 8, exceeding the boiling point of liquid nitrogen. The metallic $\sigma$-bonding bands play a crucial role in boosting the strengths of EPC. We mainly present the results for $n$=7 and 8. The calculation methods and results for H$_n$-B$_2$C$_6$ ($n$=1-6) are given in Appendix A and B, respectively.

\section{Results and discussions}

\begin{figure}[b]
\centering
\refstepcounter{figure}
\addtocounter{figure}{-1}
\includegraphics[width=8.6cm]{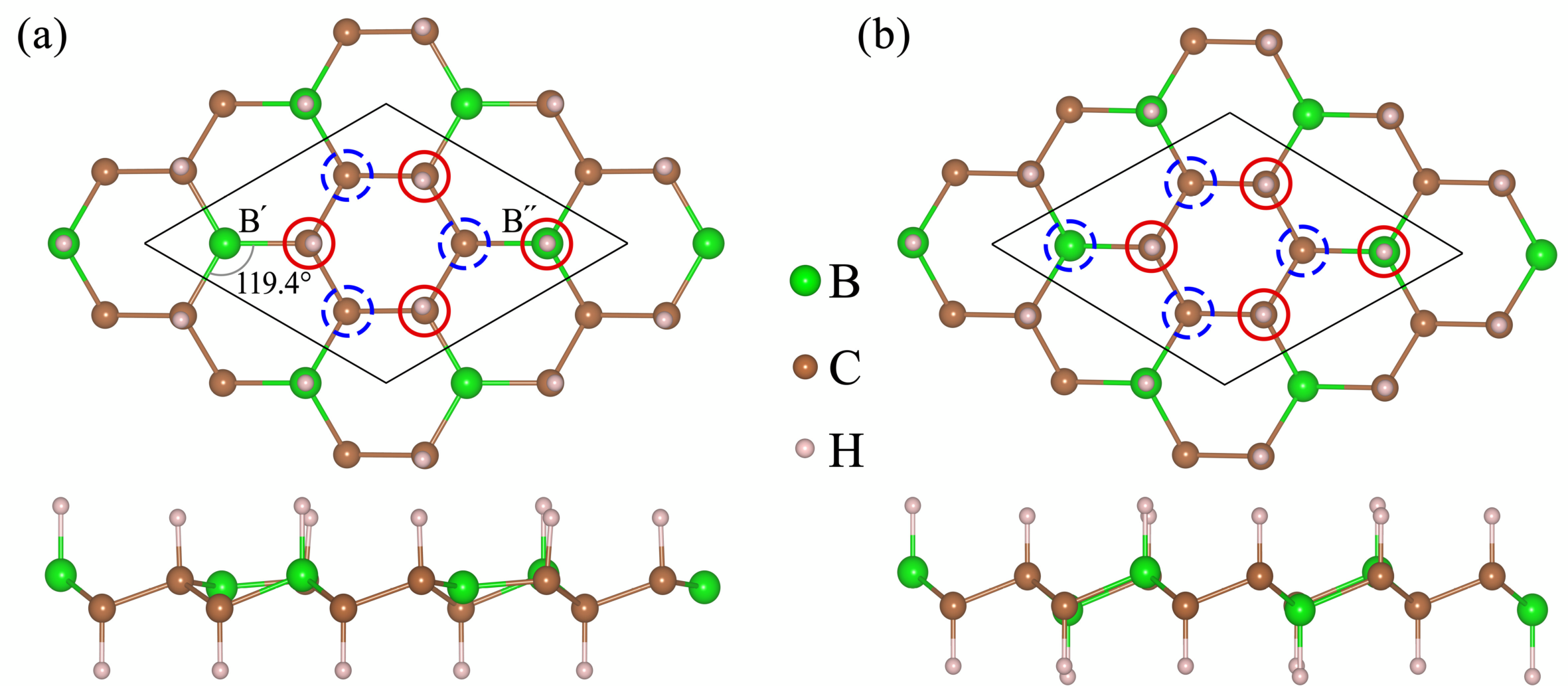}
\caption{Crystal structures of H$_7$-B$_2$C$_6$ (a) and H$_8$-B$_2$C$_6$ (b). In the top views, solid red and dashed blue circles denote hydrogen atoms adsorbed above and below the plane, respectively. Boron, carbon, and hydrogen atoms are shown in green, brown, and pink, respectively. The black lines denote the unit cell. Inequivalent boron atoms in H$_7$-B$_2$C$_6$ are labelled as B$'$ and B$''$.}
\label{fig:Stru}
\end{figure}

According to previous results \cite{Ding-JPCC113,Chuang-Nanotechnology22}, the adsorbed hydrogen atoms at the nearest neighbor sites are located on opposite sides of the sheet. The most stable crystal structures of partially hydrogenated H$_7$-B$_2$C$_6$ and fully hydrogenated H$_8$-B$_2$C$_6$ are shown in Fig.~\ref{fig:Stru}. In H$_7$-B$_2$C$_6$, B$'$ and B$''$ stand for boron atoms without and with hydrogenation. B$'$ remains in a $sp^2$-hybridized environment, whereas the C-B$'$-C bond angle marked in Fig.~\ref{fig:Stru}(a) is 119.4$^\circ$, quite close to 120$^\circ$. Hydrogenation gives rise to the formation of B-H and C-H bonds, which weakens and elongates the original planar bonding network of BC$_3$ monolayer, driving the system away from an ideal planar geometry. For example, the average B-C/C-C bond lengths are 1.615/1.544 {\AA} in H$_7$-B$_2$C$_6$ and 1.672/1.531 {\AA} in H$_8$-B$_2$C$_6$, which exhibit evident increase with respect to 1.564/1.421 {\AA} in BC$_3$ monolayer.
Consequently, the pronounced out-of-plane corrugation develops and stabilizes a $sp^3$-hybridized configuration. The optimized in-plane lattice constants are 5.232 {\AA} for H$_7$-B$_2$C$_6$ and 5.175 {\AA} for H$_8$-B$_2$C$_6$. The thicknesses of the monolayer structure, represented by maximum vertical H-H separations along the ${\bf c}$ axis, are 2.918 {\AA} and 3.116 {\AA} in H$_7$-B$_2$C$_6$ and H$_8$-B$_2$C$_6$, respectively. The larger in-plane lattice constant and smaller thickness of H$_7$-B$_2$C$_6$ are closely related to the $sp^2$-bonding nature of B$'$.

\begin{figure}[t]
\centering
\refstepcounter{figure}
\addtocounter{figure}{-1}
\includegraphics[width=8.6cm]{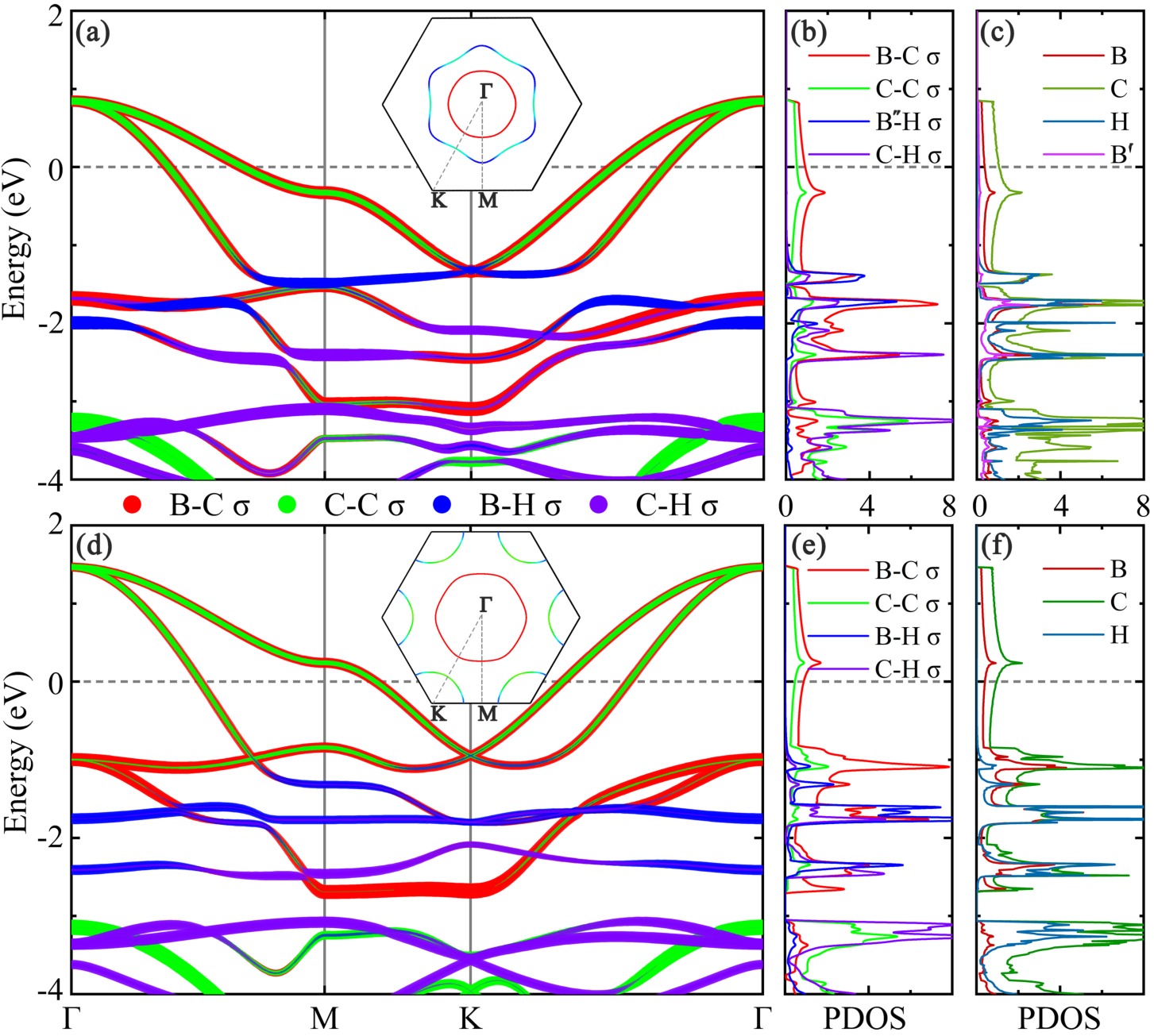}
\caption{Electronic structures of H$_7$-B$_2$C$_6$ and H$_8$-B$_2$C$_6$. (a) and (d) $\sigma$ orbital-resolved band structures, with red, green, blue, and purple colors representing the weights of B-C, C-C, B-H, and C-H $\sigma$ orbitals. Line width is proportional to the orbital weight. The Fermi level is set to zero. Fermi surfaces, weighted by the Fermi velocity are shown as insets. (b) and (e) $\sigma$-orbitals projected DOS. (c) and (f) Partial DOS with atomic resolution.}
\label{fig:band}
\end{figure}

Figure \ref{fig:band} presents the electronic structures of H$_7$-B$_2$C$_6$ and H$_8$-B$_2$C$_6$. As revealed by the band structures, there are two bands across the Fermi level both in H$_7$-B$_2$C$_6$ [Fig. \ref{fig:band}(a)] and H$_8$-B$_2$C$_6$ [Fig. \ref{fig:band}(d)], the metallic characteristics are consistent with previous results \cite{Ding-JPCC113,Chuang-Nanotechnology22}. Compared with insulating $sp^3$-bonded graphane \cite{Savini-PRL105,Sofo-PRB75}, the metallicity of H$_8$-B$_2$C$_6$ can be naturally explained by the electron deficiency of boron. For H$_7$-B$_2$C$_6$, the valence band maximum at $\Gamma$ point is about 0.85 eV above the Fermi level, which shifts upward to 1.46 eV in H$_8$-B$_2$C$_6$. This indicates that hydrogen adsorption acts as hole doping, reflecting the fact that hydrogen extracts electrons from BC$_3$ sheet, in accordance with the elongation of B-C and C-C bond lengths after hydrogen adsorption. In H$_7$-B$_2$C$_6$, two partially filled bands form two Fermi surfaces centered around the $\Gamma$ point [see inset in Fig. \ref{fig:band}(a)]. Owing to hole doping induced by hydrogenation, Fermi surface near the $\Gamma$ point expands, while another Fermi surface changes from surrounding the $\Gamma$ point to encircling the $K$ point at the corner of the Brillouin zone [inset in Fig. \ref{fig:band}(d)]. The orbital-projected band structures and DOS reveal that the electronic states near the Fermi level mainly originate from the B-C and C-C $\sigma$-bonding orbitals [Fig. \ref{fig:band}(b) and Fig. \ref{fig:band}(e)]. According to the partial DOS, electronic states associated with hydrogen and B$'$ atoms are primarily below -1.0 eV, and have negligible contribution to $N(0)$ [Fig. \ref{fig:band}(c) and Fig. \ref{fig:band}(f)]. As a result, $N(0)$ reaches 1.46 states/eV/cell in H$_7$-B$_2$C$_6$ and 1.38 states/eV/cell in H$_8$-B$_2$C$_6$, about 40.38\% and 32.69\% larger than that of 10\% hole-doped graphane \cite{Savini-PRL105}.

\begin{figure}[htbp]
\centering
\refstepcounter{figure}
\addtocounter{figure}{-1}
\includegraphics[width=8.6cm]{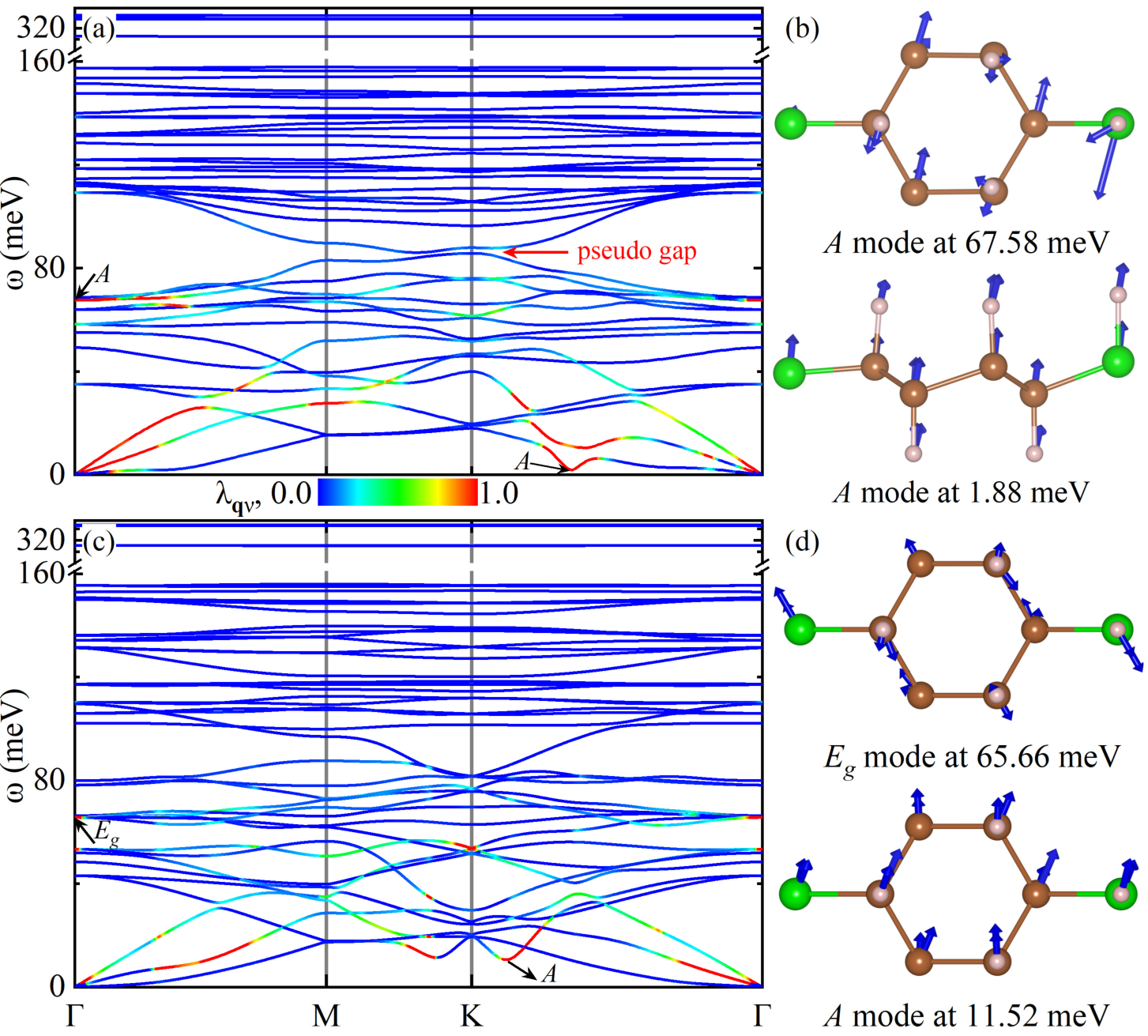}
\caption{Phonon spectra of H$_7$-B$_2$C$_6$ (a) and H$_8$-B$_2$C$_6$ (c). The color scale represents the magnitude of $\lambda_{\mathbf{q}\nu}$ at each phonon wavevector and mode. (b) and (d) Strongly coupled vibrational modes highlighted by black arrows in (a) and (c), with blue arrows indicating the direction and relative amplitude of atomic displacements.}
\label{fig:phonon}
\end{figure}

Since the metallization of $\sigma$-bonding bands is a strong signature of potential high-$T_c$ superconductivity \cite{Gao-Physics44}, such as in MgB$_2$ \cite{Nagamatsu-Nature410,An-PRL86,Kong-PRB64,Choi-Nature418}, hole-doped LiBC-related compounds \cite{Rosner-PRL88,Gao-PRB91,Gao-PRB101}, hole-doped graphane \cite{Savini-PRL105}, and B-C clathrates \cite{Zhu-SA6,Wang-PRB103,Zhang-PRB105,Gai-PRB105,Geng-JACS145,Li-PRB109},
it is thus quite interesting to know whether there exists high-$T_c$ superconductivity in H$_7$-B$_2$C$_6$ and H$_8$-B$_2$C$_6$.
%To answer this question, ***.
Figure~\ref{fig:phonon} shows the phonon spectra of H$_7$-B$_2$C$_6$ and H$_8$-B$_2$C$_6$, colored by the EPC strength $\lambda_{\mathbf{q}\nu}$ with wavevector and mode resolution. The absence of imaginary phonon frequencies confirms the dynamical stability of both systems. There is a pseudo frequency gap around 86 meV in H$_7$-B$_2$C$_6$. For H$_8$-B$_2$C$_6$, low-frequency phonon hardening smears out this gap. Moreover, phonons below the frequency gap dominate the contribution to EPC. To elucidate the vibrational modes that responsible for EPC, two representative strongly coupled phonon modes are selected at the $\Gamma$ point and along the K-$\Gamma$ path. In H$_7$-B$_2$C$_6$, the presence of an $sp^2$-hybridized B$'$ atom lowers the structural symmetry, giving rise to nondegenerate modes, for example, an in-plane $A$ mode at 67.58 meV and an out-of-plane $A$  mode at 1.88 meV [Fig. \ref{fig:phonon}(c)]. Although frequency of the second $A$ mode approaches zero and has the potential to trigger a structural phase transition, the frequency remains positive within the harmonic approximation. If anharmonic effects are considered, such strongly coupled phonons may undergo further hardening, just as anharmonic effects consistently stiffen low-frequency phonons \cite{Errea-Nature532,Errea-Nature578}. In contrast, the higher symmetry of H$_8$-B$_2$C$_6$ leads to a doubly degenerate in-plane bond-stretching $E_g$ mode and an out-of-plane $A$ mode [Fig. \ref{fig:phonon}(d)].

\begin{figure}[htbp]
\centering
\refstepcounter{figure}
\addtocounter{figure}{-1}
\includegraphics[width=8.6cm]{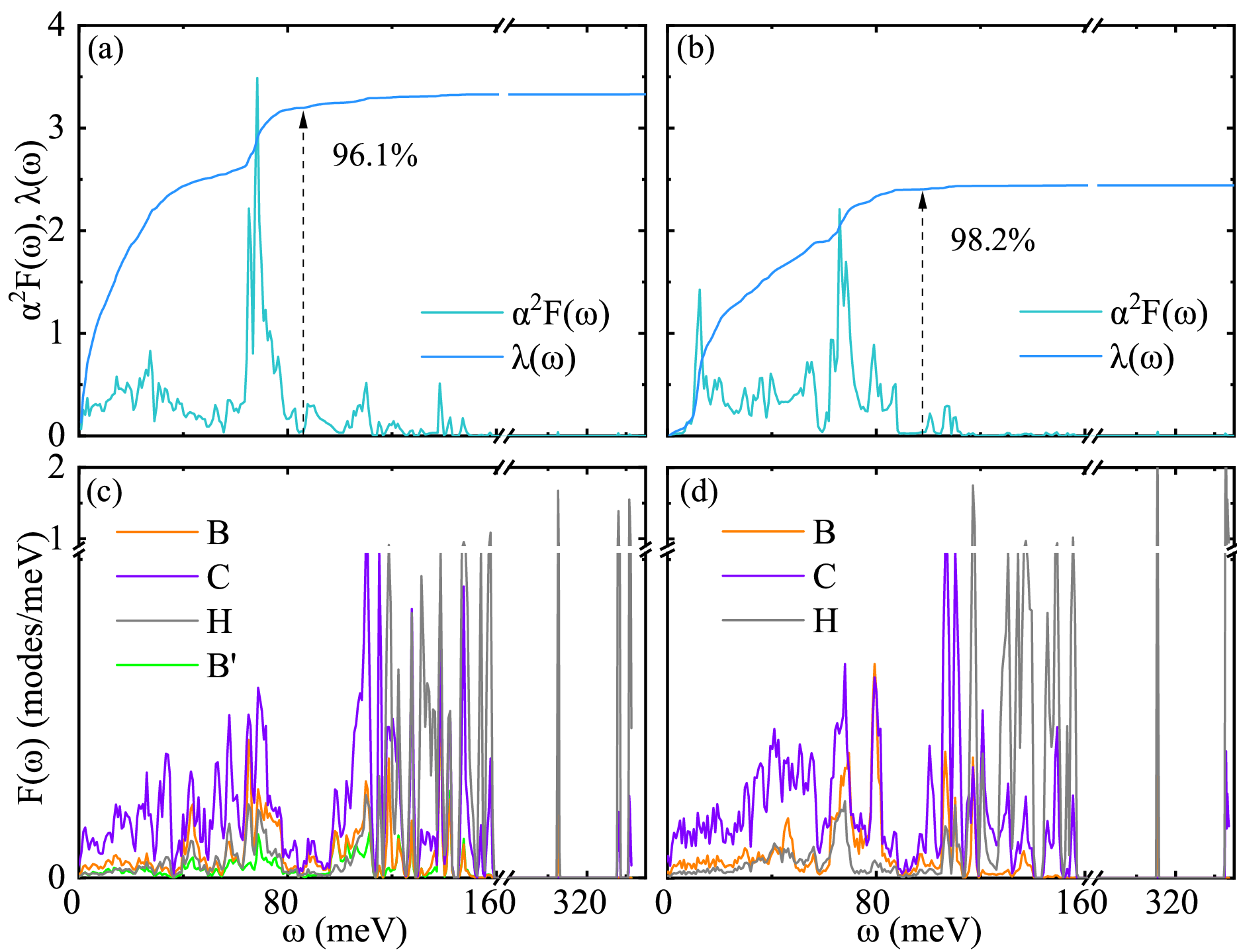}
\caption{(a) and (b) Eliashberg function $\alpha^2F(\omega)$ and accumulated $\lambda(\omega)$ of H$_7$-BC$_3$ and H$_8$-BC$_3$, respectively. Black arrows indicate the contribution of phonons below the pseudo gap to $\lambda$. (c) and (d) Projected phonon DOS $F(\omega)$. }
\label{fig:PHDOS}
\end{figure}

The total and projected phonon DOS obtained via the quasi-harmonic approximation are shown in Fig. \ref{fig:PHDOS}. Below the pseudo frequency gaps, phonon modes originate from the vibrations of boron and carbon atoms, which contribute 96.1\% and 98.2\% to the total $\lambda$ in H$_7$-B$_2$C$_6$ and H$_8$-B$_2$C$_6$, according to the accumulated $\lambda(\omega)$ obtained by $2\int_0^\omega\alpha^2F(\omega')/\omega'd\omega'$ [Fig. \ref{fig:PHDOS}(a) and Fig. \ref{fig:PHDOS}(b)]. This agrees with the distribution of $\lambda_{\mathbf{q}\nu}$ given in Fig. \ref{fig:phonon}(a) and Fig. \ref{fig:phonon}(c). Hydrogen-associated phonons govern the frequency above 100 meV [Fig. \ref{fig:PHDOS}(c) and Fig. \ref{fig:PHDOS}(d)], which have negligible proportion in EPC, reflecting that the electronic states of hydrogen are far away from the Fermi level. By integrating $\alpha^2F(\omega)$, the EPC constants $\lambda$ are calculated to be 3.33 and 2.44 in H$_7$-B$_2$C$_6$ and H$_8$-B$_2$C$_6$, about 2.29 and 1.68 times that of 10\% hole-doped graphane, respectively. The larger EPC can be attributed to the heavily hole doping, which release the bonding force of $\sigma$ bonds and cause significant phonon softening. Consequently, the logarithmic average phonon frequencies $\omega$$_{\log}$, defined by $\omega_{\log}=\exp\left[\frac{2}{\lambda}\int\alpha^2F(\omega)/\omega\ln\omega d\omega\right]$, are found to be 13.88 and 25.28 meV, evidently smaller that in 10\% hole-doped graphane. Please refer Table~\ref{tab:Tc} in Appendix B for details of other metallic phases.

\begin{figure}[htbp]
\centering
\refstepcounter{figure}
\addtocounter{figure}{-1}
\includegraphics[width=8.6cm]{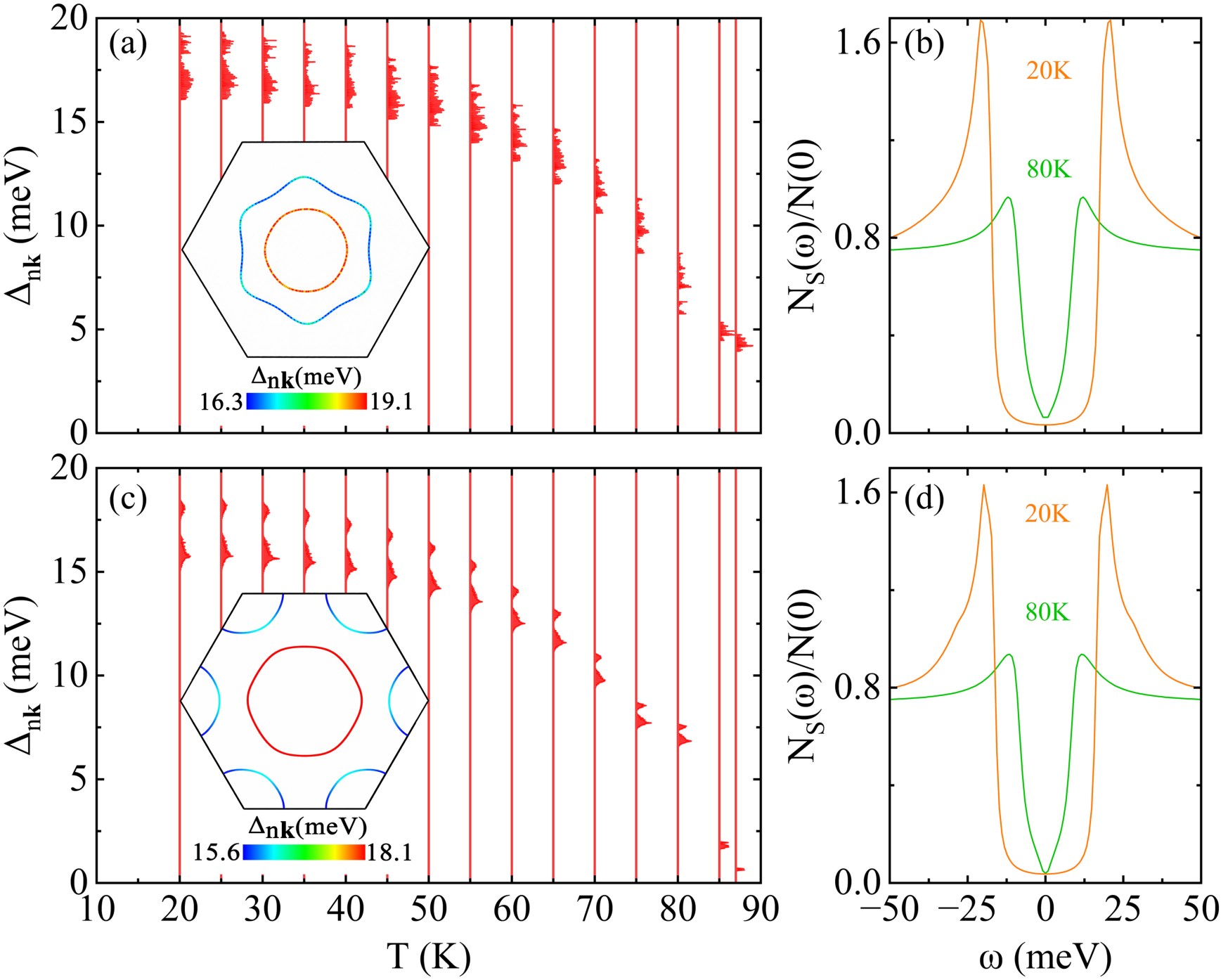}
\caption{Temperature dependence of the gap values $\Delta_{n\mathbf{k}}$ on the Fermi surfaces at different temperatures for H$_7$-B$_2$C$_6$ (a) and H$_8$-B$_2$C$_6$ (c). Insets show the distribution of superconducting gap $\Delta_{n\mathbf{k}}$ on the Fermi surfaces at 20 K. (b) and (d) Normalized quasiparticle DOS in the superconducting state at at 20 K and 80 K, respectively.}
\label{fig:Tc}
\end{figure}

Through solving the anisotropic Eliashberg equations \cite{Choi-PhysC385,Margine-PRB87}, we obtain the temperature-dependent distribution of the superconducting gap $\Delta_{n\mathbf{k}}$ and the quasiparticle DOS for H$_7$-B$_2$C$_6$ and H$_8$-B$_2$C$_6$, as shown in Fig.~\ref{fig:Tc}. The superconducting gaps group together, suggesting a single-gap feature [Fig.~\ref{fig:Tc}(a) and Fig.~\ref{fig:Tc}(c)], which is also confirmed by quasiparticle DOS at 20 K and 80 K [Fig.~\ref{fig:Tc}(b) and Fig.~\ref{fig:Tc}(d)]. The gaps $\Delta_{n\mathbf{k}}$ on inner pocket surrounding the $\Gamma$ point have larger values. At 20 K, the average superconducting gap $\Delta_{n\mathbf{k}}^{\mathrm{ave}}$ is 17.3 meV and 16.3 meV for H$_7$-B$_2$C$_6$ and H$_8$-B$_2$C$_6$.
The gap anisotropy ratio $\Delta_{n\mathbf{k}}^{\rm Aniso}$, defined as $(\Delta_{n\mathbf{k}}^{\max}-\Delta_{n\mathbf{k}}^{\min})/\Delta_{n\mathbf{k}}^{\mathrm{ave}}$, is 16.2\% for H$_7$-B$_2$C$_6$ and 15.3\% for H$_8$-B$_2$C$_6$. The highest temperature at which a nonvanishing $\Delta_{n\mathbf{k}}$ is obtained defines the superconducting transition temperature. It is interesting that both systems have the same $T_c$ of 87 K, quite close to the value predicted in hole-doped graphane \cite{Savini-PRL105}, but much higher than that predicted in metallic $ABC$-stacked BC$_3$ \cite{Ribeiro-PRB69}.

\section{Conclusion}

In summary, this study demonstrates that hydrogen serves as a potent design element for inducing and enhancing high-temperature superconductivity in two-dimensional materials. Through systematic first-principles investigations of hydrogenated BC$_3$ monolayer, we find that hydrogenation not only stabilizes a chair-like $sp^3$-hybridized structure but also effectively dopes holes into the system, leading to $\sigma$-band metallization and a substantial increase in the density of states at the Fermi level. The strong coupling between these metallic $\sigma$ electrons and low-frequency phonon modes results in a significant electron-phonon interaction. By solving the anisotropic Eliashberg equations, it is revealed that specific hydrogen coverages, particularly in H$_7$-B$_2$C$_6$ and H$_8$-B$_2$C$_6$, can achieve a superconducting transition temperature up to 87 K, surpassing the liquid-nitrogen boiling point. This work establishes hydrogenated BC$_3$ as a promising candidate for high-$T_c$ superconductivity, offering a feasible material platform that bypasses the need for extreme pressures or complex correlated electron systems. It highlights a strategic approach to designing practical, high-temperature superconductors by combining dimensional confinement, tailored electronic structures, and hydrogen-mediated electron-lattice coupling, thereby advancing the prospects for liquid-nitrogen-temperature superconducting applications.

\begin{acknowledgments}

This work was supported by the National Key R\&D Program of China (Grants No. 2024YFA1408601 and No. 2023YFA1406201), Zhejiang Provincial Natural Science Foundation of China (Grants No. LMS26A040005 and No. LMS26A040006), the National Natural Science Foundation of China (Grants No. 12434009 and No. 12274255), Program for Science and Technology Innovation Team in Zhejiang (Grant No. 2021R01004), and Ningbo Young Scientific and Technological Innovation Leaders Program (Grant No. 2023QL016). F.M was also supported by the BNU Tang Scholar.

\end{acknowledgments}

\appendix

\section{Calculation Methods}

In our computations, we utilize the plane-wave basis set approach as implemented in the QUANTUM ESPRESSO package \cite{Giannozzi-JPCM21}. A vacuum layer of 20 {\AA} is added along the {\bf c} axis to avoid the coupling between adjacent sheets. The exchange-correlation potentials are described using the generalized gradient approximation (GGA) based on the Perdew-Burke-Ernzerhof (PBE) functional \cite{Perdew-PRL77}. For modeling electron-ion interactions, we adopt the optimized norm-conserving Vanderbilt pseudopotentials \cite{Hamann-PRB88,Schlipf-CPC196}. Through convergence tests, we set the kinetic energy cutoff and charge density cutoff to 80 Ry and 320 Ry, respectively. The charge densities are calculated self-consistently on an unshifted mesh of 24$\times$24$\times$1 points, using a Methfessel-Paxton smearing of 0.02 Ry \cite{Methfessel-PRB40}. The dynamical matrices and perturbation potentials are computed on $\Gamma$-centered meshes of 6$\times$6$\times$1 for H$_n$-B$_2$C$_6$ ($n$=1, 3, 5, and 7) and 8$\times$8$\times$1 for H$_8$-B$_2$C$_6$, using density-functional perturbation theory \cite{Baroni-RMP73}.

The parameters used to construct maximally localized Wannier functions (MLWFs) \cite{Pizzi-JPCM32} are as follows. Twenty MLWFs are employed to describe the corresponding band structures, including $sp^2$-hybridized $\sigma$-like states and $p_z$ orbital for B/C atom without hydrogen adsorption, and $sp^3$-hybridized orbitals for hydrogenated B/C atoms. The {\bf k} meshes are set to 6$\times$6$\times$1 for H$_n$-B$_2$C$_6$ ($n$=1, 3, 5, and 7) and 8$\times$8$\times$1 for H$_8$-B$_2$C$_6$, respectively. Utilizing the EPW package \cite{Lee-npjCM9}, the Brillouin zone is sampled with three different fine {\bf k} meshes: $120 \times 120 \times 1$, $180 \times 180 \times 1$, and $240 \times 240 \times 1$, meanwhile employing a $60 \times 60 \times 1$ {\bf q} mesh, to assess the convergence of the EPC constant $\lambda$. After convergence test, the smearing widths of the electron and phonon Dirac $\delta$ functions are set to 30 meV and 0.5 meV, respectively. When solving the anisotropic Migdal-Eliashberg equations, a 180$\times$180$\times$1 mesh for electrons and a 60$\times$60$\times$1 mesh for phonons are adopted. The Matsubara frequency is truncated at about 3.5 eV, which is roughly ten times the highest phonon frequency.

\section{Results of H$_n$-B$_2$C$_6$ ($n$=1-6)}

\begin{figure}[tbh]
\centering\includegraphics[width=8.6cm]{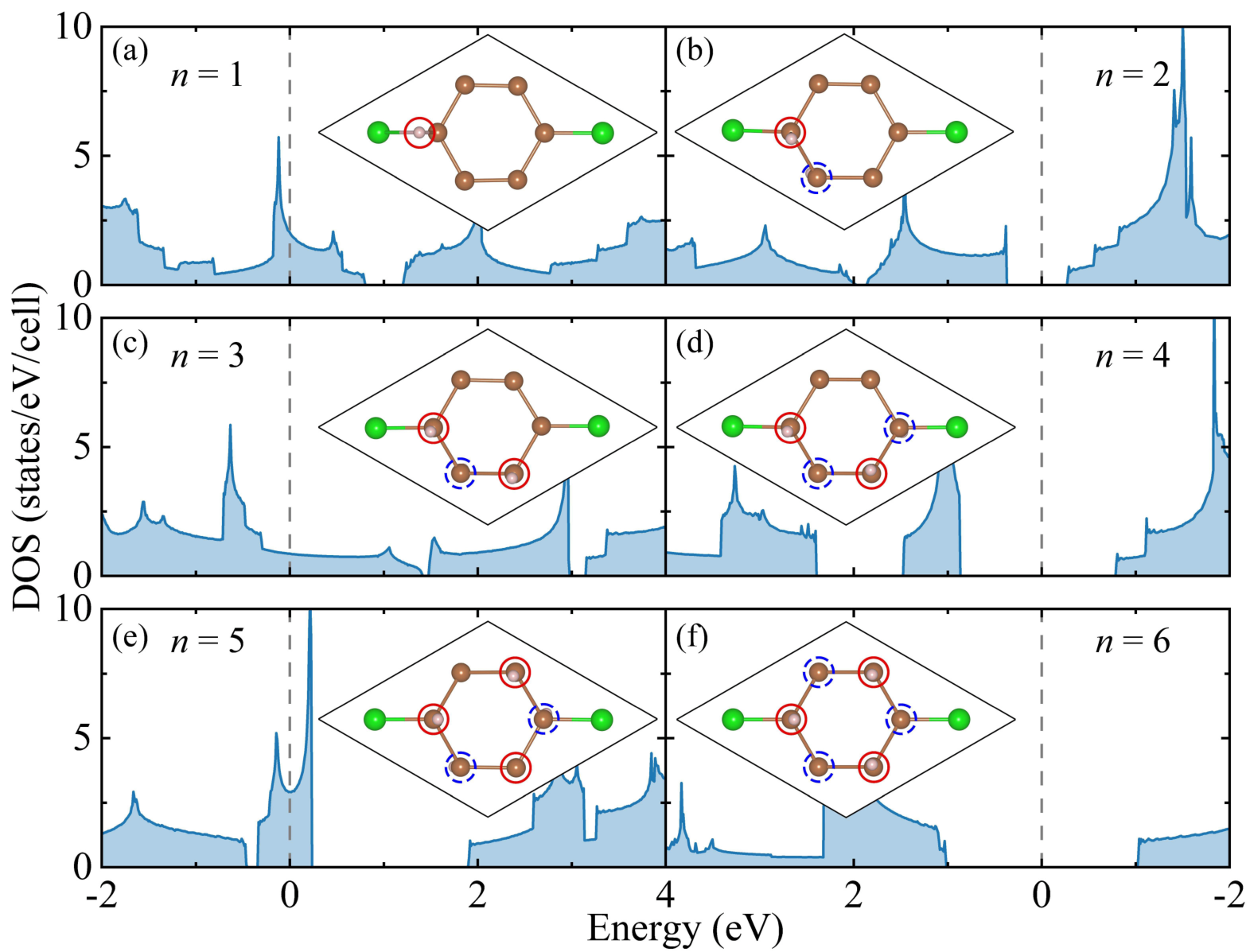}
\caption{Ground-state structures and DOS for H$_n$-B$_2$C$_6$ ($n$=1-6).}
\label{fig:DOS}
\end{figure}

According to previous first-principles calculations \cite{Chuang-Nanotechnology22}, the ground-state configurations of hydrogenated BC$_3$ sheets exhibit distinct structural trends as a function of hydrogen coverage. At very low coverages, hydrogen atoms preferentially adsorb on carbon sites rather than boron. The most favorable arrangement places hydrogen atoms on opposite sides of the sheet when they attach to neighboring carbon atoms. From H$_1$-B$_2$C$_6$ to H$_6$-B$_2$C$_6$, hydrogen atoms tend to occupy carbon sites within the same hexagonal ring, progressively filling adjacent carbon sites with alternating side adsorption. With increasing coverage up to H$_6$-B$_2$C$_6$, a fully carbon-terminated structure is obtained. Beyond this coverage, hydrogen begins to adsorb on boron atoms, eventually leading to a fully saturated H$_8$-B$_2$C$_6$ structure. The ground-state crystal structures of H$_n$-B$_2$C$_6$ ($n$=1-6) are shown as insets in Fig. \ref{fig:DOS}.
It is clear that H$_n$-B$_2$C$_6$ are semiconducting for $n$=2, 4, 6 [Fig. \ref{fig:DOS}].

\begin{table}[b]
\centering
\caption{Crystal structure, electronic structure, EPC, and $T_c$ for metallic H$_n$-B$_2$C$_6$.
$a$ and $b$ represent the in-plane lattice constants, with $\gamma$ denoting the angle between them.
$N(0)$ is the total DOS at the Fermi level. $\omega$$_{\log}$ and $\sqrt{\langle \omega^2 \rangle}$ are two characteristic phonon frequencies, defined in McMillan-Allen-Dynes (MAD) formula \cite{Allen-PRB12}.
$T_{c,0.1}^{\mathrm{MAD}}$ and $T_{c,0.1}^{\mathrm{Aniso}}$ stand for the $T_c$ evaluated by the MAD formula and the anisotropic Eliashberg equations, respectively, with the Coulomb pseudopotential $\mu^*$ set to 0.1. The units for length, angle, DOS, energy, frequency, and $T_c$ are ${\AA}$, $^\circ$, states/eV/cell, eV, meV, and K.}
\setlength{\tabcolsep}{3pt}
\begin{tabular}{c c c c c c c c c c}
\toprule
$n$
& $a $
& $b $
& $\gamma$
& $N(0)$
& $\lambda$
& $\omega_{\log}$
& $\sqrt{\langle \omega^2 \rangle} $
& $T_{c,0.1}^{\mathrm{MAD}}$
& $T_{c,0.1}^{\mathrm{Aniso}}$ \\
\midrule
1  & 5.151  & 5.151 & 119.1 & 1.98 & 3.07 & 7.23  & 32.49  & 22.7 & 32 \\
%2  & 5.141  & 5.187 & 119.7 & 0.00 & 0.00 & 0.00  &  0.00  & 0.0  & -- \\
3  & 5.218  & 5.218 & 120.3 & 0.84 & 0.00 & 65.08 & 110.17 & 0.0  & -- \\
%4  & 5.214  & 5.214 & 119.5 & 0.00 & 0.00 & 0.00  &  0.00  & 0.0  & -- \\
5  & 5.273  & 5.260 & 120.1 & 2.89 & 0.33 & 19.14  & 68.23  & 0.2  & -- \\
%6  & 5.289  & 5.289 & 120.0 & 0.00 & 0.00 & 0.00  &  0.00  & 0.0  & -- \\
7  & 5.231  & 5.231 & 120.0 & 1.46 & 3.33 & 13.88  & 41.16  & 47.9 & 87 \\
8  & 5.175  & 5.175 & 120.0 & 1.38 & 2.44 & 25.28  & 43.12  & 63.5 & 87 \\
\bottomrule
\end{tabular}
\label{tab:Tc}
\end{table}

Table~\ref{tab:Tc} summarizes the structural parameters, electronic properties, EPC, and possible superconductivity of metallic phases of H$_n$-B$_2$C$_6$ ($n$=1, 3, 5, 7, and 8). With increasing hydrogen coverage, the in-plane lattice constant varies nonmonotonically due to competing lattice expansion and local distortions. H$_n$-B$_2$C$_6$ gradually evolves from $sp^2$ to $sp^3$ hybridization, and at high coverage the resulting $sp^3$ network causes the Fermi-level states to be dominated by $\sigma$ electrons. The $N(0)$ and EPC constant $\lambda$ also show a pronounced nonmonotonic trend: H$_1$-B$_2$C$_6$ exhibit $\lambda > 3.0$ and sizable $T_c$, whereas H$_3$-B$_2$C$_6$ has vanishing EPC. Notably, although H$_5$-B$_2$C$_6$ has a relatively large $N(0)$, detailed analysis shows that $N(0)$ stems from isolated $\pi$ orbitals, and weak EPC is naturally obtained. For H$_7$-B$_2$C$_6$ and H$_8$-B$_2$C$_6$, sizable $\lambda$ and moderate phonon frequencies yield high $T_c$ up to 87 K. Overall, the dependence of $T_c$ on hydrogen coverage is non-monotonic.

\end{document}